\documentstyle[numreferences]{kapproc}
\runningtitle{COALGEBRAS AND COHOMOLOGY}
\begin{opening}
\title{COALGEBRAS, COCOMPOSITIONS AND COHOMOLOGY
\thanks{The contribution to Third International Conference
on Non Associative Algebra and Its Applications, Oviedo, Spain,
July 1993. Proceedings edited by Santos Gonz\'alez are published by Kluwer
Academic Publishers in a series Mathematics and its Applications.}}
\subtitle{}
\author{Zbigniew Oziewicz\thanks{Permanent address: University of Wroc{\l}aw,
Po
\institute{Universidad Nacional Autonoma de
M\'exico\\oziewicz@redvax1.dgsca.una
\author{Eugen Paal}\institute{University of Tallin, Department of
Mathematics,\\5 Ehitajate tee, Tallinn EE0108, Estonia\\paal@ce.ttu.ee}
\author{Jerzy R\'o\.za\'nski\thanks{All authors partially supported by the
Polis
Scientific Research (KBN), Grant \# 2 2419 92 03.
}}\institute{University of Wroc{\l}aw,\\
plac Maxa Borna 9, 50204 Wroc\l aw, Poland\\rozanski@plwruw11.bitnet}
\end{opening}

\newtheorem{(a,b)-derivation}{Definition}
\newtheorem{cocomutator}[(a,b)-derivation]{Definition}

\newtheorem{ja}[(a,b)-derivation]{Definition}
\newtheorem{factor derivation}{Corollary}
\newtheorem{l1}{Lemma}
\newtheorem{Eugen}[l1]{Lemma}
\newtheorem{complex}[l1]{Lemma}
\newtheorem{l2}[l1]{Lemma}
\newtheorem{Q}[l1]{Lemma}
\newtheorem{invariance}[l1]{Lemma}
\newtheorem{l3}[l1]{Lemma}

\begin{document}

\newcommand{\ds}{\displaystyle}
\newcommand{\un}{\underline}
\newcommand{\be}{\begin{equation}}
\newcommand{\ee}{\end{equation}}
\newcommand{\ba}{\begin{array}}
\newcommand{\ea}{\end{array}}
\newcommand{\ben}{\begin{enumerate}}
\newcommand{\een}{\end{enumerate}}
\def\Zint{{Z \kern -.45em Z}}
\def\Nint{I\!\!N}
\def\Cint{I\!\!\!\!C}
\newcommand{\ra}{\longrightarrow}
\newcommand{\id}{\mbox{id}}
\newcommand{\der}{\mbox{der}}
\newcommand{\ring}{\mbox{ring}}
\newcommand{\mod}{\mbox{mod}}
\newcommand{\alg}{\mbox{alg}}
\newcommand{\im}{\mbox{im}}
\newcommand{\End}{\mbox{End}}
\newcommand{\gen}{\mbox{gen}}
\newcommand{\lin}{\mbox{lin}}
\newcommand{\grade}{\mbox{grade}}
\newcommand{\eid}{{\em id}}
\newcommand{\eder}{{\em der}}
\newcommand{\ering}{{\em ring}}
\newcommand{\emod}{{\em mod}}
\newcommand{\ealg}{{\em alg}}
\newcommand{\eim}{{\em im}}
\newcommand{\eEnd}{{\em End}}
\newcommand{\egen}{{\em gen}}
\newcommand{\elin}{{\em lin}}
\newcommand{\edeg}{{\em deg}}
\newcommand{\edim}{{\em dim}}
\newcommand{\egrade}{{\em grade}}
\def\ss{\begin{tabular}{c}}
\def\kk{\end{tabular}}

\begin{abstract}
The {\em (co)homology theory} of $n$-ary (co)compositions is a functor
associating to $n$-ary (co)composition a complex. We present unified approach
to the cohomology theory of coassociative and Lie
coalgebras and for $2n$-ary cocompositions. This approach points to a possible
ge\-ne\-ralization.


\end{abstract}

\section*{Foreword}
Let $k$ be a field of characteristic $0,$
$L$ be a finite dimensional $k$-vector space
and $L^*\equiv\lin(L,k)$ be the dual $k$-space of $L.$
Let the tensor product $\otimes $ means ${\otimes }_{k},$  $TL\equiv
L^{\otimes}$ be the tensor algebra generated by $k\oplus L$ and let $I\lhd TL$
be a two-sided ideal.
The linear map $\triangle\in\lin(L,L^{\otimes n})$ is said to be the
$n$-ary noncoassociative cocomposition on $L.$
In other words, {\em a noncoassociative $n$-ary cocomposition}, coalgebra for
$n
$\triangle $ is a (n,1)-tensor over $L,$
\be\triangle\in L^{\otimes n}\otimes L^*.\ee
By $D_\triangle\in\der(TL)$ we are denoting the
$\Nint$-homogeneous (skew)derivation of the tensor algebra such that
$$D_\triangle|k\equiv 0\quad\mbox{and}\quad D_\triangle|L\equiv \triangle.$$
The {\em cohomology theory} of $n$-ary cocompositions is a functor
$$\{L,\triangle\}\longmapsto\{TL/I,D_\triangle/I\},$$
associating to $n$-ary cocomposition (coalgebra), $\{L,\triangle\},$ a complex
$\{TL/I,D_\triangle/I\}$.
Results proven for finite-dimensional cocompositions (coalgebras)
give the valid results for compositions (algebras) by categorical duality.

Cohomology and homology theory of associative algebras was
formulated by Hochschild in 1945. One year after, Claude
Chevalley together with Samuel Eilenberg,
inspired by \'{E}lie Cartan implicite considerations,
gave the precise definition of the cohomology and
homology groups of the Lie algebras (published in 1948).
Independently the (co)homology
theory of Lie algebras was defined by Jean-Luis
Koszul (Th\'{e}se 1950). Both theories, Hochschild theory
and Chevalley - Eilenberg - Koszul theory,
are treated in most of textbooks as separate theories. A unified
axiomatic treatment of both theories with the help of the derived
functors Tor (derived from $\otimes$) and Ext (derived from Hom)
is presented by Henri Cartan
and Samuel Eilenberg in the monograph {\em Homological Algebra}
(1956).

The Cartan - Eilenberg axiomatic (co)homology theory
does not make clear, why (co)homology theory should make
sense for two varieties of algebras, associative and Lie, only?
Already in 1948 Eilenberg pose the problem of an association of a cohomology
groups to {\em any} nonassociative algebra. Yamaguti in 1963
formulated the cohomology theory for algebras
introduced by Malcev in
1955 on the basis of the analytic loops. Cohomology of the Lie triple systems
has been elaborated by Harris in 1961 and by Yamaguti (1957-1969).
The (co)homology theory was extended to the graded algebras and graded Lie
triples (Tilgner 1977) and make
sense for Lie
$\Zint_2$-graded algebras and Lie multi-graded algebras (Bani and Tripathy
1984)
One can look for a complete list of algebras for which a
(co)homology theory make sense. To do this we ask what
is special about the associative and Lie algebras which allows to
formulate for them a (co)homology theory? The present notes
are motivated by this question.

We present the unified approach to the cohomology
theory of $n$-ary cocompositions and in particular the unified approach
to the cohomology theory of associative and Lie coalgebras.

Slight modification of presented approach can be applied for the non-Lie Malcev
algebras, however this is beyond of the scope of this short note.

\section*{Graded derivations}
Denote $\wedge\equiv\otimes/I.$
\begin{(a,b)-derivation} Let $\alpha,\,\beta\in\ealg(TL/I).$ An
$(\alpha,\beta)$-{\em derivation} of the factor algebra $TL/I$ is a linear map
$\delta\in\elin(TL/I)$ that satisfies Leibnitz's rule
$$\delta\circ\wedge\equiv\wedge\circ(\delta\otimes \alpha+\beta\otimes
\delta):\quad(TL/I)\otimes(TL/I)\ra TL/I.$$
The vector $k$-space of all ($\alpha,\beta$)-derivations of $TL/I$ is denoted
by
$\eder_{\alpha,\beta}(TL/I).$
\end{(a,b)-derivation}

We will determine the conditions on the linear maps
$\alpha,\,\beta,\,\delta\in\lin(TL/I)$
which for $p\in\Nint$ assure the implication
$$\delta\in\der_{\alpha,\beta}\Longrightarrow\delta^p\in\der_{\alpha^p,\beta^p}.
The definition imply that for each $p\in\Nint$ we have
$$\delta^p\circ\wedge=\wedge\circ(\delta\otimes\alpha+\beta\otimes \delta)^p.$$
Therefore the condition we are looking for is
$$(\delta\otimes\alpha+\beta\otimes\delta)^p
=\delta^p\otimes\alpha^p+\beta^p\otimes \delta^p.$$
We can solve this condition for the $\Nint$-homogeneous case. Let
$D\in\der_{A,B}(TL)$ be $\Nint$-homogeneous derivation of
grade$\,D\equiv|D|\equ
n-1.$ The definition imply
$$|A|=|B|=0,\quad A1=B1=1\quad\mbox{and}\quad D1=0.$$ The equation
$$(D\otimes A+B\otimes D)^p=D^p\otimes A^p+B^p\otimes D^p$$ for $p>1$ and for
every homogeneous derivation $D$ imply that
$$A_\xi|L=\xi\cdot\id_L\quad\mbox{and}\quad B_\zeta|L=\zeta\cdot\id_L,$$ where
$\xi,\,\zeta\in k.$
\begin{l1} The implication
$$D\in \eder_{\xi,\zeta}(TL) \Longrightarrow D^p\in \eder_{\xi^p,\zeta^p}(TL)$$
holds iff $\xi\zeta=0$ or if $\xi\zeta\ne 0$ and
$(\zeta/\xi)^{|D|}$ is the nontrivial $p^{th}$ root of unity,
$\zeta^{p|D|}=\xi^{p|D|},\;\zeta \ne \xi$.\\ In particular:
\begin{eqnarray*}
D\in \eder_{\xi,\zeta}(TL) \Longrightarrow D^2\in \eder_{\xi^2,\zeta^2}(TL)
&\qquad\mbox{iff}&\qquad \qquad \xi\zeta(\xi^{|D|}+\zeta^{|D|})=0.\\
D\in \eder_{\xi,\zeta}(TL) \Longrightarrow D^3\in \eder_{\xi^3,\zeta^3}(TL)
&\qquad\mbox{iff}&\qquad
\xi\zeta(\xi^{2|D|}+\xi^{|D|}\zeta^{|D|}+\zeta^{2|D|})=
\end{eqnarray*}
\end{l1}
The same statment holds true for any $\Nint$-graded factor algebra
$TL/I$. The algebra $\{TL/I, B_\zeta\}$ is $\Zint_p$-graded for
$\zeta^p=1.$
The implication of Lemma 1 for $p=3$ has been related by Kerner (1992) to
the cubic root of the translations.

The $\Nint$-homogeneous components of the tensor algebra $TL$ are denoted by
$$T^nL\equiv L^{\otimes n}\equiv \{t\in TL;\;\grade\,t=n\}.$$
If $|D|=+1$ and $\xi=1$ then we have the recurrent formula
\begin{eqnarray*}
D^2|L^{\otimes n}&=&(D^2|L^{\otimes
(n-1)})\otimes\id_L+\zeta^{2(n-1)}\id_{L^{\otimes
(n-1)}}\otimes(D^2|L)\\&+&\zeta^{n-1}(\zeta
+1)(D|L^{\otimes(n-1)})\otimes(D|L).
\end{eqnarray*}
This prove the lemma
\begin{Eugen}
Let $D\in\eder_{\zeta}(TL),$ $|D|=+1,$ then
$$D^2=0\quad\Longleftrightarrow\quad D^2|L=0\quad\mbox{and}\quad D^2|L^{\otimes
\end{Eugen}

\section*{Factor derivations}
Let $\pi$ be the epimorphism: $TL\ra TL/I.$ A linear map
$A\in\lin_k(TL)$ factors to $A/I\in\lin(TL/I)$ if and only if
$$AI\subset I\quad\mbox{and then}\quad
\pi\circ A=(A/I)\circ\pi.$$
The map $$\omega:\quad (TL)\otimes (TL)\ra TL$$ factors to the
map $\omega/I:(TL/I)\otimes(TL/I)\ra TL/I,$ iff
$$\omega(I\otimes TL+TL\otimes I)\subset I,\quad\mbox{and
then}\quad\pi\circ\ome
If the factor map $A/I$ exists then exists the factor map $(A\circ\otimes)/I$
and $$(A\circ\otimes)/I=(A/I)\circ(\otimes/I):\quad (TL/I)\otimes(TL/I)\ra
TL/I.
Let $A$ and $B$ be linear maps in $\lin(TL).$ If exist the factor maps, $A/I$
an
$B/I,$ then exists the factor maps $(A\circ B)/I,$ $\{\otimes\circ(A\otimes
B)\}
$$(A\circ B)/I=A/I\circ B/I,$$
$$\{\otimes\circ(A\otimes B)\}/I=(\otimes/I)\circ(A/I\otimes B/I):\quad
(TL/I)\otimes(TL/I)\ra TL/I.$$
\begin{factor derivation}
The derivation $D\in\eder_{A,B}(TL)$ factors to the $(A/I,B/I)$-derivation
$D/I$
of the factor algebra $D/I\in\eder_{A/I,B/I}(TL/I)$ if and only if
$A,\,B$ and $D$ preserve the ideal $I$.
\end{factor derivation}

\section*{Homogeneous ideals}
The ideal $I\lhd TL$ is said to be {\em homogeneous} if
$$I=\sum \{I\cap(L^{\otimes n})\},\quad I\cap(k\oplus L)=0.$$
For homogeneous ideal the factor algebra $TL/I$ is $\Nint$-graded,
$\grade\circ\pi=\grade,$ and $\grade\,\wedge=0.$
Let the map $D\in\lin(TL)$ be $\Nint$-homogeneous with $|D|=|D/I|.$ In this
case $|A|=|A/I|=0.$

Let $A$ and $B$ be zero-grade algebra maps, $A,\,B\in\alg_0(TL),$ such that
$$A|L\equiv\xi\cdot\id_L\quad\mbox{and}\quad B|L\equiv\zeta\cdot\id_L.$$
These algebra maps preserve {\em homogeneous} ideals and therefore factors to
the factor-algebra maps $A/I,\,B/I\in\alg(TL/I).$

The cocomposition $\triangle\in\lin(L,L^{\otimes n})$ determine the unique
$(\xi,\zeta)$-derivation $D_\triangle$ of the tensor algebra,
$$\triangle\hookrightarrow\;D_\triangle\in\der_{\xi,\zeta}(TL),$$
such that $|D_\triangle|=n-1$ and $$D_\triangle|k\equiv
0\quad\mbox{and}\quad D_\triangle|L\equiv\triangle.$$
If the ideal $I\lhd TL$ is homogeneous then the linear map
$\triangle\in\lin(L,L^{\otimes n})$ determine the unique
$(\xi,\zeta)$-derivation
$(D_\triangle/I)\in\der_{\xi,\zeta}(TL/I)$ of the factor
algebra iff \be D_\triangle I\subset I.\ee

\section*{Cocomutative n-ary cocomposition}
\begin{cocomutator} Let $I\lhd TL$ be an ideal. The $n$-ary cocomposition
$\tria
L^{\otimes n}\otimes L^*$ is said to be $I$-{\em cocomutative} if
$$({\em im}\,\triangle)\cap I=0.$$
\end{cocomutator}
The condition of definition determine the variety of $I$-cocomutative
noncoassociative cocompositions (coalgebras).

Let $P\in\End(L^{\otimes n})$ be idempotent such that
\be I\equiv\gen\{\im\, P\}\,\lhd\, TL.\ee
The $(n,1)$-tensor $P\circ\triangle$ over $L$ is said to be $P$-{\em
skew}-cocomutator of the cocomposition (coalgebra) $\{L,\triangle\}.$

A $P$-(skew) cocommutative $n$-ary cocomposition (coalgebra) is the triple
$\{L,P,\triangle \}$ where
the (n,1)-tensor $\triangle $ and the idempotent (n,n)-tensor
$P$ satisfy the conditon $P\circ \triangle =0,$
$\im\,\triangle \,\subset \,\ker P.$

\section*{Jacobi condition for $n$-ary cocomposition}
Let $$\xi\zeta(\xi^{(n-1)}+\zeta^{(n-1)})=0.$$ Then
$$(D_\triangle/I)^2\in\der_{\xi^2,\zeta^2}(TL/I)$$ and therefore
$$(D_\triangle/I)^2=0\quad\Longleftrightarrow\quad (D_\triangle/I)^2|L=0.$$
We have $$(D/I)^2\circ\pi=\pi\circ D^2.$$ Therefore
\be
(D_\triangle/I)^2=0\;\Longleftrightarrow\;\fbox{im$(D_\triangle\circ\triangle)\s
I\cap L^{\otimes(2n-1)}$}.\ee
The framed condition is said to be the Jacobi condition for $n$-ary
cocomposition $\triangle.$
The more general Jacobi condition could be
$$D^m/I=(D/I)^m=0\Longleftrightarrow \im\, (D^m)\subset I.$$
We proved the following
\begin{complex}
Let $\triangle$ be $n$-ary co-composition on $L$ with a derivation
$D_\triangle\in\der_{\xi,\zeta}(TL),$ $|D_\triangle|=n-1.$ Let $I$ be {\em
homogeneous} ideal in the tensor algebra $TL.$ The necessary and sufficient
cond
the map $$\{L,\triangle\}\longmapsto \{TL/I,D_\triangle/I\},$$
{\em define} the cohomology theory for $n$-ary co-composition $\triangle,$ are
\begin{description}
\item{(i)} $\xi^{n-1}+\zeta^{n-1}=0\quad(\Longrightarrow
(D_\triangle)^2\in\eder
\item{(ii)} Invariance $D_\triangle I\subset I\qquad(\Leftrightarrow
\exists\;D_\triangle/I\in\eder\;\mbox{and}
\;\exists\;(D_\triangle)^2/I=(D_\triangle/I)^2\in\eder),$
\item{(iii)} $I$-cocomutativity (\eim$\,\triangle)\cap I=0\in L^{\otimes n}$,
\item{(iv)} The Jacobi condition: $\eim(D_\triangle\circ\triangle)\subset I\cap
L^{\otimes(2n-1)}\quad(\Leftrightarrow (D_\triangle/I)^2=0).$
\end{description}
\end{complex}
The $(\xi,\zeta)$-derivation $D_\triangle\in\der_{\xi,\zeta}(TL)$ can be viewed
$D~\equiv ~\{D|L^{\otimes n}\}$,
\be D_\triangle|L^{\otimes n}=
\sum_{k=0}^{n-1}{\zeta}^k \id_k\otimes \triangle\otimes
\xi^{n-k-1}\id_{n-k-1},\
where $D_\triangle|L\equiv\triangle$ and
$$\xi\zeta(\xi^{(n-1)}+\zeta^{(n-1)})=0
The above formula tells
that a derivation $D$ is determined by a parameters $\{\xi,\zeta\}$
and values of $D$ on
generating space $L$, $\triangle\equiv D|L\hookrightarrow D$. In the case of
the
tensor algebra $TL,$ these values may be chosen arbitrarily.
In particular $D=0 \Longleftrightarrow D|L=0$.
If $D^p$ is a derivation then $D^p=0$ is
equivalent to $D^p|L=0$.

Let $P\in\End (L^{\otimes n})$ and $Q\in\End (L^{\otimes(2n-1)})$ be
idempotents
\be I\equiv\gen\{\im P\}\quad\mbox{and}\quad\ker Q\equiv I\cap L^{\otimes
(2n-1)}.\ee
In terms of the operators $P,\,Q$ and
$\triangle$ we have altogether three compatibility conditions which assure the
cohomology theory for $n$-ary co-compositions.

The existence of the factor-derivation $D_\triangle I\subset
I\quad\Leftrightarrow$
\be \fbox{$Q\circ D_\triangle\circ P=0$}:\quad L^{\otimes n}\ra L^{\otimes
(2n-1
A $P$-(skew)-cocommutativity
\be \fbox{$P\circ\triangle=0$}:\quad L\ra L^{\otimes n}.\ee
The Jacobi condition
\be \fbox{$Q\circ D_\triangle\circ\triangle=0$}:\quad L\ra L^{\otimes
(2n-1)}.\e
In equations (7-9) $D_\triangle$ is given by the formula (5).

One can consider the set of equations (7-9) from the two points of view:
\begin{itemize}
\item Let the ideal $I$ be given. This means that the operators $P$ and $Q$ are
given. Then the above set of three equations determine (possible empty) variety
cocompositions satisfying identities. We will show in the next section that in
this way one can {\em define} the coassociative coalgebras and Lie coalgebras
as
examples of the varieties of coalgebras possesing the cohomology theory.
\item Let the $n$-ary cocomposition $\{L,\triangle\}$ be given. For
example we can take (binary) noncoassociative coalgebra. Then the
above set of equations determine (possible empty) set of (quadratic) ideals in
t
tensor algebra $TL,$ each of which determine the cohomology theory for the
given
noncoassociative coalgebra $\{L,\triangle\}.$
\end{itemize}

\section*{Coassociative $2n$-ary cocompositions.}
The choice $$I\equiv 0,\quad
P\equiv 0\quad\mbox{and}\quad Q\equiv \id_{L^{\otimes 3}},$$ define the
$2n$-ary
coassociative} cocompositions, which are analogous to the coassociative
coalgebras, and posess the cohomology theory.
For this choice we are left with the Jacobi condition only.

For example the $4$-ary
cocomposition is said to be coassociative if the (7,1)-tensor over $L,$
$$(\triangle\otimes\id_{L^{\otimes
3}}-\id_L\otimes\triangle\otimes\id_{L^{\otimes 2}}+\id_{L^{\otimes
2}}\otimes\triangle\otimes\id_L-\id_{L^{\otimes
3}}\otimes\triangle)\circ\triang
is the zero tensor. This tensor is the analogy of the coassociator for the
binary cocompositions.

For $|D_\triangle|=1,$ $$D_\triangle|(L\otimes L)=\triangle\otimes
\xi{id}_L+\ze
\triangle,$$
and $(D_\triangle)^2$ is a derivation iff $\xi\zeta(\xi+\zeta)=0.$
Three solutions leads to
\begin{eqnarray*}
\mbox{if}\;\zeta=0\phantom{+\xi}&\qquad&(D_\triangle)^2|L=(\triangle\otimes\id_L
\mbox{if}\;\xi=0\phantom{+\xi}&\qquad&(D_\triangle)^2|L=(\id_L\otimes\triangle)\
\mbox{if}\;\xi+\zeta=0&\qquad&(D_\triangle)^2|L=(\triangle\otimes\id_L-\id_L\oti
\end{eqnarray*}
The last expression is the {\sl coassociator} of a coalgebra $\{L,\triangle\}.$
Therefore for $I\equiv 0$ the Jacobi condition, $(D_\triangle)^2=0,$ determine
the variety of coassociative coalgebras
$$(\triangle\otimes\id_L-\id_L\otimes\triangle)\circ\triangle=0:\quad L\ra
L^{\otimes 3}.$$

The {\it coassociativity}
is equivalent to the condition $(D_\triangle)^2=0$ and in this way
every {\it coassociative} co-algebra $\{L,\triangle\}$
have the (scalar-valued) {\it complex}
$\{TL,D_\triangle\}.$ The cohomology groups of the complex
$\{TL,D_\triangle\}$ are {\it by
definition} the cohomology groups (with the scalar coeficients) of the
{\it coassociative} co-algebra $\{L,\triangle\}.$
This is the scalar-valued Hochschild
cohomology. The presented scheme
allows the generalizations for not coassociative coalgebras.

The Lie algebras are skew-symmetric and in this case
$I$ is the quadratic ideal generated by the symmetric tensors.
We are going to show that this ideal, the formula (10) below, is
preserved by the $(\xi,\zeta)$-derivations
if and only if $\xi+\zeta=0$ and that this statement is
$\triangle$-independent. For this we need some notations
related to the representation of the braid and permutation groups in
the tensor algebra $TL$.

\section*{Braid and permutation groups}
A convenient way to describe the braid group $B_n$ is
by means of the two generators $\tau$ and $c$
(Artin 1926, Coxeter and Moser 1957, page 63)
$$B_n\equiv\{\tau,c|c^n=(c\circ \tau )^{n-1}=(\tau \circ c)^{n-1}\}.$$
Let $c$ be represented by the cyclic isomorphism
$c\in\End (L^{\otimes n}),$
such that $c^n|L^{\otimes n}=id,$ $$ c(v_1\otimes v_2\otimes \dots \otimes v_n)
(-)^{n-1}\cdot (v_n\otimes \dots v_{n-1}),$$
and let $$N|L^{\otimes n}\equiv c+c^2+\dots +c^n.$$
Therefore $(c-id)\circ N=0.$

Let $\tau$ be represented by the twist isomorphism, ${\tau}^2=\mbox{id}_{TL},$
o
$$\tau (v_{1} \otimes v_{2} \otimes v_{3} \otimes \dots ) \equiv v_{2}
\otimes v_{1} \otimes v_{3} \otimes \dots $$
For $n\leq 3,\; \tau \circ N=N\circ\tau.$

\begin{l2}
For any comultiplication $\triangle\in L\otimes L\otimes L^*$ the identities
hold
\begin{eqnarray*}
c\circ (\triangle \otimes id_L)&=&(id_L\otimes \triangle)\circ\tau\\
c^2\circ(id_L\otimes \triangle)&=&(\triangle \otimes id_L)\circ \tau.
\end{eqnarray*}
\end{l2}

\section*{Lie coalgebras}
The paper by
Michaelis (1980) is the general reference about Lie coalgebras.
Let \be P\equiv(\id_{L^{\otimes 2}}+\tau)\in\End(L\otimes
L)\quad\mbox{and}\quad
I\equiv\gen\{\im\, P\}.\ee
\begin{Q}
$$I\equiv\egen\{\eim\,P\}\quad\Longrightarrow\quad
I\cap L^{\otimes 3}=\ker\{(\id_{L^{\otimes 3}}-\tau)\circ N\}.$$
\end{Q}
In this case
$$Q=N\circ(\id_{L^{\otimes 3}}-\tau)=(\id_{L^{\otimes 3}}-\tau)\circ
N\in\End(L^{\otimes 3}).$$
Note that $\frac{1}{6}Q$ is the projector on $L\wedge L\wedge L\equiv
L^{\wedge
3}.$
\begin{invariance}
The ideal (10) is preserved by the $(\xi,-\xi)$-derivation $D_\triangle,$
$D_\triangle I\subset I,$ for {\em every} comultiplication $\triangle.$
\end{invariance}
\noindent{\bf Proof.} From Lemma 4 we have the identity
$$N\circ D_\triangle\circ P=0\quad\mbox{for each}\quad\triangle.$$
\hfill$\Box$\\
More generaly, we conjecture that $D_\triangle$-stable quadratic ideals,
$\forall\:\triangle,$ needs to
satisfy the Yang-Baxter equation, as it is in the case of the
braided Lie algebras invented by Gurevich.

For the ideal (10) we are left with the two conditions (8-9) only.
\begin{ja}
The (3,1)-tensor $J\equiv J_\triangle$ over $L$
$$J\equiv
Q\circ(\triangle\otimes\id_L-\id_L\otimes\triangle)\circ\triangle:\quad L\ra
L^{\otimes 3},$$ is said to be the {\em coJacobiator} of coalgebra
$\{L,\triangle\}.$
\end{ja}
The ideal (10) determine the variety of the co-skew-symmetric coalgebras
\be P\circ\triangle=0.\ee
\begin{l3}
The {\it co-Jacobiator} $J$ of a co-skew-commutative coalgebras (11) possess
the
alternative forms:
\begin{eqnarray*}
J &=& +4N\circ (\triangle \otimes id_L)\circ \triangle\\
  &=& -4N\circ (id_L\otimes\triangle)\circ \triangle.
\end{eqnarray*} \end{l3}
Therefore the ideal (10) determine the variety of Lie coalgebras, satisfying
the
$\tau\circ~\triangle\equiv-\triangle$ and $J_\triangle\equiv 0,$ {\bf as the
variety of
coalgebras possessing the cohomology theory}.


\begin{thebibliography}{99}
\bibitem{Chevalley} Chevalley Claude and Samuel Eilenberg: 1948,
     `Cohomology of Lie groups and Lie algebras',
     {\em Trans. AMS} {\bf 63} 85--124
\bibitem{Coxeter} Coxeter H.S.M. and W.O.J. Moser: 1957,
    {\em Generators and relations for discrete groups}, (Springer-Verlag
    Berlin, G{\"o}ttingen)
\bibitem{Eilenberg} Eilenberg Samuel: 1948, `Extensions of general algebras',
{\
Polon. Math.}, (Rocznik Polskiego Towarzystwa Matematycznego) {\bf 21} (1)
125-134
\bibitem{Harris} Harris B.: 1961, `Cohomology of Lie triple systems and Lie
alge
      with involutions',\\ {\em Transactions of American Mathematical Society}
{
        148-162
\bibitem{Hochschild} Hochschild G.: 1945, `On the cohomology groups of an
      associative algebra',\\ {\em Annals of Mathematics} {\bf 46} 58-67
\bibitem{Kerner} Kerner Richard: 1992, `$\Zint_3$-graded algebras and cubic
      root of the supersymmetry translations',
     {\em Journal of Mathematical Physics} {\bf 33} (1) 403--411
\bibitem{Koszul} Koszul Jean-Louis: 1950, `Homologie et cohomologie des
    alg\'ebras de Lie', Th\'ese,\\
    {\em Bull. Soc. Math. France} {\bf 78} 1--63
\bibitem{Michaelis} Michaelis Walter: 1980, `Lie coalgebras', {\em Advances in
Mathematics}
{\bf 38} 1--54
\bibitem{Mitra} Mitra Bani and K.C. Tripathy: 1984, `The cohomology of the
      generalized Lie algebras',\\ {\em J. Math. Phys.} {\bf 25} 2550-2556
\bibitem{Tilgner} Tilgner Hans: 1977, `A graded generalization of Lie triples',
{\em Journal of Algebra}, {\bf 47} 190-196
\bibitem{Yamaguti57} Yamaguti Kiyosi: 1957-58, `On the Lie triple system and
its
       generalization', {\em Journal Sci. Hiroshima University}, {\bf A21}
155-1
\bibitem{Yamaguti63} Yamaguti Kiyosi: 1963, `On the theory of Malcev algebras',
{\em Kumamoto Journal Sci.} {\bf A6} (1) 9-45
\bibitem{Yamaguti69} Yamaguti Kiyosi: 1969, `On cohomology groups of general
Lie
               systems', {\em Kumamoto J. Sci.}, {\bf A8} (4) 135-146
\end{thebibliography}
\end{document}